# Analyzing the Resource Utilization of Lambda Functions on Mobile Devices: Case Studies on Kotlin and Swift

Chibundom U. Ejimuda, Gaston Longhitano, Reza Rawassizadeh
cejimuda, gastonl, rezar @bu.edu
Department of Computer Science, Metropolitan College, Boston University

**Abstract**—With billions of smartphones in use globally, the daily time spent on these devices contributes significantly to overall electricity consumption. Given this scale, even minor reductions in smartphone power use could result in substantial energy savings. This study explores the impact of Lambda functions on resource consumption in mobile programming. While Lambda functions are known for enhancing code readability and conciseness, their use does not add to the functional capabilities of a programming language. Our research investigates the implications of using Lambda functions in terms of battery utilization, memory usage, and execution time compared to equivalent code structures without Lambda functions. Our findings reveal that Lambda functions impose a considerable resource overhead on mobile devices without offering additional functionalities. All codes are available at: https://github.com/iamchibu/lambdafunctions

In recent decades, we have observed that smartphones outnumber personal computers [1], emphasizing their widespread accessibility. It is estimated that 7.33 billion smartphones are operated worldwide [2]. A recent study [3] indicates Android is and iPhone users spend, on average, 3:15' on their phones daily. This surge in availability positions smartphones as powerful tools for various applications. As these devices predominantly rely on battery power, considering their daily electricity consumption becomes a concern for electricity production.

In the US, around 75% of the battery of smartphones is utilized daily from morning to 10 pm [4]. Budget smartphones usually have about 2500 mAh, and high-end ones have 6000 mAh battery capacity, the average smartphone battery could be assumed to be 4100 mAh [5]. Based on these data, we estimate that *3,075 mAh* of electricity is used daily per smartphone. Assuming there are 7.33 billion smartphones available, we can estimate **$2.253975 \times 10^{13}$** *mAh* of electricity spent on smartphones worldwide. Assuming a defacto-standard charger of five volts is used for charging, the total electricity spent on smartphone charges daily worldwide will be approximately **112,698,750 kWh.** This estimation shows the enormous amount of electricity used daily by smartphones. We believe any effort to reduce the electricity usage of smartphones, such as energy-efficient software development practices, will have an enormous environmental impact and reduction in $CO_2$ emissions. This observation can motivate us to explore more energy-efficient programming and software development practices.

Any digital advancement is associated with a massive amount of electricity usage and, thus $CO_2$ emissions. For example, recently, after the massive commercialization of Large Language Models, we observed another spike in electricity consumption [6, 7, 8].

Most of the smartphone market is operated by two major operating systems: Android (uses Kotlin and Java programming languages) and iPhone (uses Swift programming language). Both Kotlin and Swift support the use of lambda function (a.k.a. Anonymous function or closure in Swift).

Lambda functions have their roots in lambda calculus, a formal system in mathematical logic and computer science for expressing computation based on function abstraction and application. The concept was introduced by mathematician Alonzo Church [9], long before the advent of modern programming languages. Lambda functions in modern programming platforms introduce a more concise and expressive syntax [10], and some features for efficient coding, including lazy evaluation, improved code readability, and the mitigation of code duplication [11]. However, it is still possible to benefit from all platform capabilities without Lambda functions.

This research investigates the impact of Lambda functions on mobile devices. In particular, we study the use versus not using the lambda function on memory consumption, battery utilization, and execution time on mobile devices. We develop experimental codes in both Kotlin (for Android devices) and Swift (for iPhones) programming languages, incorporating Lambda functions in one variant and omitting them in the other.

Our findings reveal notable differences in resource utilization, between using Lambda versus not using Lambda variants. We measure and report execution time, memory, and battery use. Our experiments show that while Lambda functions offer concise code, they may incur significant resource overhead on resources. Therefore, by avoiding the use of the Lambda functions in mobile programming, we believe an enormous amount of electricity will be preserved daily.

**METHODOLOGY**

To study the impact of the lambda function and its features, we chose the two mobile phone operating system platforms, Android and iOS, and studied the use of the Lambda function in their programming languages. First, we list different features that the use of lambda function features provides in both Kotlin and Swift programming languages. Next, we describe our approach for resource usage monitoring and measurements that we implemented on both platforms.

*Lambda Function Features*

Kotlin, a modern statically typed programming language for the Java Virtual Machine, has also embraced functional programming concepts, including lambda functions. It is the standard platform used for Android development. Swift is Apple's declarative framework for building user interfaces in iOS. It supports functional programming paradigms, allowing developers to leverage lambda functions and closures for concise and expressive code. Eleven different types of operations can be done with the Lambda function in both Kotlin and Swift platforms [12, 13]. In the following, we explain them briefly.

**1) Higher-order functions:** Both Kotlin and Swift leverage higher-order functions extensively. Kotlin's support for higher-order functions allows for the encapsulation of complex operations within succinct, reusable code constructs, enhancing the expressive capacity and modularity of the programming environment. Similarly, Swift leverages higher-order functions to pass functions as parameters and return functions from other functions, resulting in more modular and reusable components.

**2) Function types:** Kotlin introduces function types, enabling the declaration of variables that can hold references to functions. This facilitates dynamic behavior and functional programming patterns. In Swift, closures represent function types, allowing developers to declare variables that hold references to closures, providing a mechanism for encapsulating functionality.

**3) Instantiating a function/closure type:** Both languages allow for the instantiation of function/closure types, offering flexibility in defining functions on the fly. Kotlin enables the instantiation of function types, facilitating dynamic behavior and functional programming patterns. Swift allows closures to be assigned to variables, making it convenient to create dynamic functionality within components.

**4) Invoking a function/closure type instance:** Both Kotlin and Swift enable the invocation of function/closure instances based on runtime conditions, enhancing interactivity and responsiveness in applications.

**5) Inline functions/closures:** Both Kotlin's inline functions and Swift's closures optimize performance by eliminating the overhead of function calls, resulting in more efficient code execution and smoother user interfaces.

**6) Lambda/closure expression syntax:** Kotlin's lambda expressions and Swift's closure expressions provide concise and expressive syntax for defining anonymous functions, enhancing code readability and clarity.

**7) Passing trailing lambdas/closures:** Kotlin and Swift allow the passing of trailing lambdas/closures, contributing to a clean and expressive coding style and enhancing the readability of code.

**8) Implicit name of a single parameter without lambda/closure:** Kotlin implicitly uses "it" as the name for a single parameter in lambda functions, while Swift allows closures to implicitly refer to parameters by their position, simplifying syntax and making code more succinct.

**9) Returning a value from a lambda/closure expression:** Both Kotlin's lambda expressions and Swift's closure expressions can return values, adding versatility to their application in functional programming constructs.

**10) Underscore for unused variables:** Kotlin and Swift both allow the use of an underscore as a placeholder for unused variables in lambda/closure expressions, enhancing code clarity and indicating intentional unused variables.

**11) Function/closure literals with receiver:** Kotlin supports function literals with receivers, allowing the extension of existing types with new functionalities, contributing to a modular codebase. In Swift, although there's no direct support for closure literals with receivers, the concept is achieved through methods and properties on components, enabling developers to extend existing types with new functionalities for a modular and extensible codebase.

*Experiment Design*

Inspired by similar approaches to benchmark resource utilization on smartphones [14, 15] in our experiment, we established a mock task that involves summing numbers from 1 up to a specified maximum number, with iterations ranging from 10,000 to 100,000. As previously mentioned, our experiment involves implementing each of the 11 Lambda function features on both platforms, i.e., Android and iOS.

For both scenarios—utilizing Lambda Functions and not—we repeat each experiment a minimum of 10 times. Despite having ten repetitions, the presence of anomalies in some results necessitated additional iterations of certain experiments to effectively identify and eliminate outliers. Moreover, to evaluate the impact of Lambda Functions on scalability, we systematically execute our operation in a loop.

This process begins with 10,000 operations and increases in increments of 10,000, culminating in 100,000 operations, specifically following the sequence {10,000, 20,000, and 100,000}. We measure the differences between using the Lambda function versus not using the Lambda function for battery utilization, execution time, and memory usage.

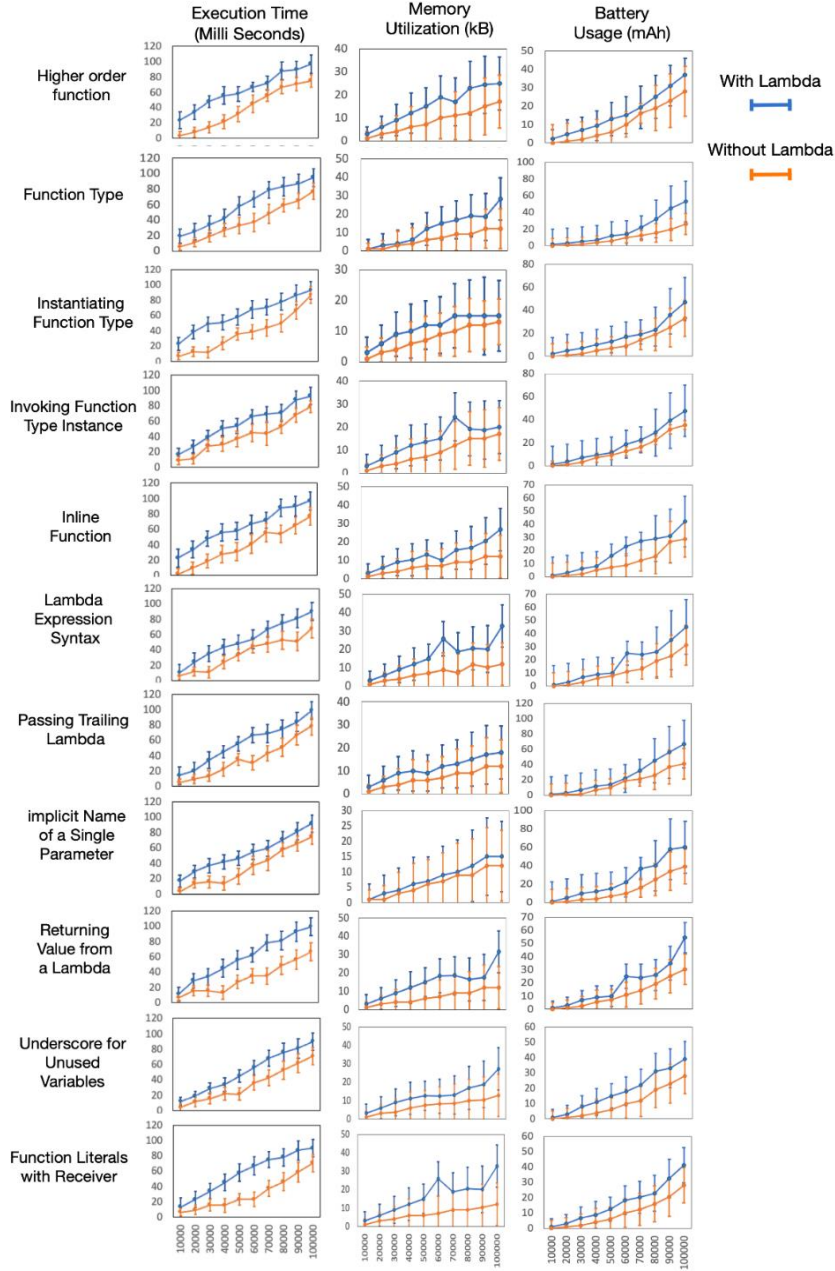

**FIGURE 1. The impact of using lambda function versus not using a Lambda Function on Android platform resource utilization. Memory is reported in kB (kilo bytes) and Battery is reported in mAh (mill ampere/hour).**

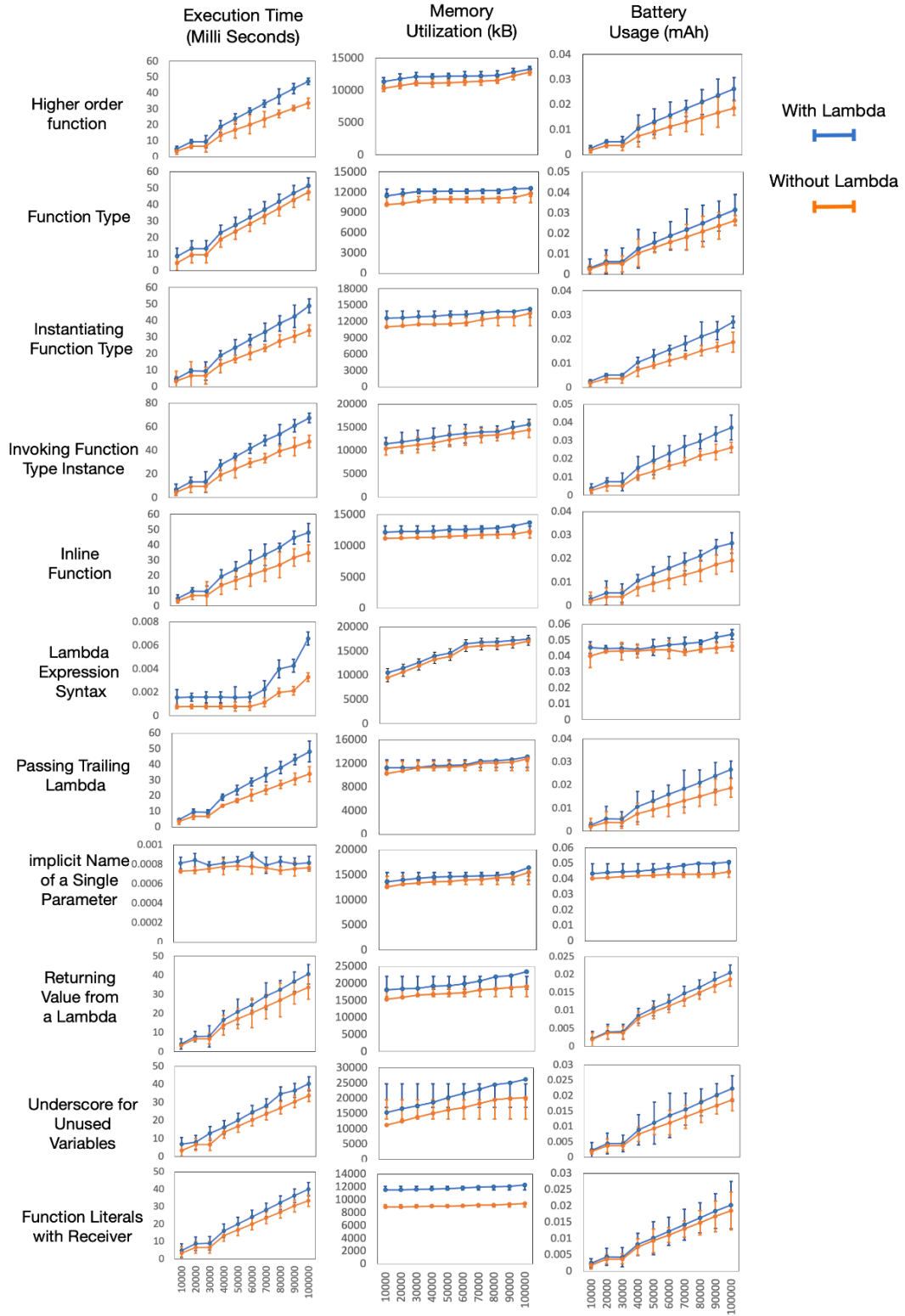

**FIGURE 2.** The impact of using lambda function versus not using a Lambda Function on iOS platform resource utilization. Memory is reported in kB (kilo bytes) and Battery is reported in mAh (mill ampere/hour).

*Experiment Setup*

For the Android platform, we have used a Samsung Galaxy A14, with Octa-core CPU (Octa-core (4x2.0 GHz Cortex-A55 & 4x2.0 GHz Cortex-A55), ARM Mali-G52 GPU, and 4GB of RAM. The iPhone that is used for our experiment is iPhone 12 Pro with an Apple A14 Bionic processor with 6-cores (2x3.1 GHz + 4x1.8 GHz) CPU, 4-core GPU, and 6 GB of RAM.

**RESULT & DICUSSIONS**

Figure 1 presents a line plot of each different form of Lambda function for the Kotlin platform. The first column reports the execution time in milliseconds, the second column shows the memory utilization in KB, and the last column shows Battery usage in mAh.

As is shown in Figure 1, by using Lambda, the function imposes 204,800 kB overhead on memory for 10,000 number summations and 409,600 kB overhead on memory for 100,000 summations. Respectively by using Lambda, the function imposes 0.5% overhead on battery utilization for 10,000 number summations and 2% overhead on battery utilization for 100,000 summations. The execution time of not using Lambda is 62.5% faster for 10,000 summation and 37.2% faster for 100,000 summation operations than using a Lambda function.

Figure 2 presents a line plot of each different form of Lambda function for the SWIFT platform. The first column reports the execution time in milliseconds, the second column shows the memory utilization in kB, and the last column shows Battery usage in mAh.

As is shown in Figure 2, by using Lambda, the function imposes 13.2% overhead on memory for 10,000 number summations and 11.1% overhead on memory for 100,000 summations. Respectively by using Lambda, the function imposes 8.9% overhead on battery utilization for 10,000 number summations and 19.9% overhead on battery utilization for 100,000 summations. The execution time of not using Lambda is 34% faster for 10,000 summation and 23.1% faster for 100,000 summation operations than using a Lambda function.

Results from both SWIFT and Kotlin platforms lead us to conclude that the performance overhead (memory, execution time, and battery) of using Lambda on both Android and iOS platforms is significant, especially in operations that include loops or require scalability.

One might argue that in an application, the performance differences between using Lambda and not using Lambda seem too insignificant to change the development policy. Nevertheless, scaling the application to the massive smartphone app markets will lead to significant changes in $CO_2$ emission by using or not using the lambda function. Therefore, we recommend not using the Lambda function for mobile development.

**CONCLUSION**

In this work, we investigate the resource use of the Lambda function on two of the most popular mobile development platforms, Android and iOS. We studied memory, battery, and execution time.

To conduct our experiments, we define a task of iterative summation of numbers, starting from 1 to 10,000. In the next iteration, it increases 10,000 numbers until it ends up with 100,000 iterations. The experiment comprised at least ten repetitions for each scenario (with and without the incorporation of Lambda Functions). We demonstrate that all eleven Lambda function operations pose an overhead on the memory and battery resources and delay the execution time.
Although its overhead might seem insignificant in a single application, considering the huge mobile application market, the $CO_2$ emission of using a lambda function is significant.